%% file: asia236-miettinen.tex
\documentclass[letterpaper]{sig-alternate-2013}

\usepackage{color}
\usepackage{url}
\usepackage{subfig}
\usepackage{xspace}
\usepackage{tabularx}
\usepackage{stfloats}
\usepackage{graphicx}
\usepackage{amsmath}
\usepackage{amssymb}

\input{macros}

\title{ConXsense -- Automated Context Classification for Context-Aware Access Control}

\numberofauthors{5}
\author{
\alignauthor Markus Miettinen
	\affaddr{Technische Universit\"at Darmstadt\\}
	\affaddr{markus.miettinen\\@trust.cased.de}
\alignauthor Stephan Heuser 
	\affaddr{Technische Universit\"at Darmstadt\\}
	\affaddr{stephan.heuser\\@trust.cased.de}
\and
\alignauthor Wiebke Kronz \\
	\affaddr{Technische Universit\"at Darmstadt}
	\affaddr{wiebke.kronz@cased.de}
\alignauthor Ahmad-Reza Sadeghi
	\affaddr{Technische Universit\"at Darmstadt}
	\affaddr{\mbox{ahmad}.sadeghi@trust.cased.de}
\alignauthor N. Asokan\\
	\affaddr{Aalto University and University of Helsinki\\}
	\affaddr{asokan@acm.org}
}

\newtheorem{definition}{Definition}

\newfont{\mycrnotice}{ptmr8t at 7pt}
\newfont{\myconfname}{ptmri8t at 7pt}
\let\crnotice\mycrnotice%
\let\confname\myconfname%

\permission{Permission to make digital or hard copies of all or part of this work for personal or classroom use is granted without fee provided that copies are not made or distributed for profit or commercial advantage and that copies bear this notice and the full citation on the first page. Copyrights for components of this work owned by others than ACM must be honored. Abstracting with credit is permitted. To copy otherwise, or republish, to post on servers or to redistribute to lists, requires prior specific permission and/or a fee. Request permissions from permissions@acm.org.}
\conferenceinfo{ASIA CCS'14,}{June 3--6, 2014, Kyoto, Japan.}
\copyrightetc{Copyright 2014 ACM \the\acmcopyr}
\crdata{978-1-4503-2800-5/14/06\ ...\$15.00.\\
http://dx.doi.org/10.1145/2590296.2590337}

\begin{document}
\maketitle
\begin{abstract}
We present \ouremphname, the first framework for context-aware access control on mobile devices based on context classification.
Previous context-aware access control systems often require users to laboriously specify detailed policies or they rely on pre-defined policies not adequately reflecting the true preferences of users. 
We present the design and implementation of a context-aware framework that uses a probabilistic approach to overcome these deficiencies. The framework utilizes context sensing and machine learning to automatically classify contexts according to their security and privacy-related properties. We apply the framework to two important smart\-phone-related use cases: protection against device misuse using a dynamic device lock and protection against sensory malware. We ground our analysis on a sociological survey examining the perceptions and concerns of users related to contextual smart\-phone security and analyze the effectiveness of our approach with real-world context data. We also demonstrate the integration of our framework with the \flaskdroid~\cite{flaskdroidUsenix} architecture for fine-grained access control enforcement on the Android platform.
\end{abstract}
\category{D.4.6}{Operating Systems}{Security and Protection}[Access controls, Invasive software]
\keywords{Mobile security; Context sensing; Privacy policies; Context-awareness}
\section{Introduction}
\label{intro}
\input{intro}


\section{Problem Description}
\label{problem}
\input{problem}

\section{Framework Description}
\label{framework}
\input{framework}
\section{User Survey }
\label{userstudy}
\input{userstudy}

\section{Use Cases}
\label{usecases}
\input{usecases}
\section{Context Model}
\label{contextmodel}
\input{contextmodel}

\section{Implementation}
\label{implementation}
\input{implementation}

\section{Evaluation}
\label{evaluation}	
\input{evaluation}

\section{Enforcement}
\label{accesscontrollayer}

\input{accesscontrol}

\section{Related Work}
\input{relatedwork}

\label{relatedwork}

\section{Conclusions and Future Work}
\label{conclusions}
\input{conclusions}



\end{document}

%% file: macros.tex
\newcommand{\ourname}{ConXsense\xspace}
\newcommand{\ouremphname}{\emph{ConXsense}\xspace}

\newcommand{\LockPatternKeyguardView}{\textsf{Lock\-Pat\-tern\-Key\-guard\-View}\xspace}
\newcommand{\Settings}{\textsf{Set\-tings}\xspace}

\newcommand{\Broadcast}{\textsf{Broad\-cast}\xspace}
\newcommand{\Intent}{\textsf{In\-tent}\xspace}
\newcommand{\PolicyServer}{\textsf{Po\-li\-cy\-Ser\-ver}\xspace}

\newcommand{\SensorManager}{\textsf{Sen\-sor\-Ma\-na\-ger}\xspace}
\newcommand{\SensorEventListeners}{\textsf{Sen\-sor\-E\-vent\-Lis\-ten\-ers}\xspace}

\newcommand{\accesscontrollayer}{\textsf{Access\ Control\ Layer}\xspace}
\newcommand{\Service}{\textsf{Service}\xspace}

\newcommand{\Lockscreen}{\textsf{Lockscreen}\xspace}
\newcommand{\AccessControlPolicy}{\textsf{Ac\-cess Con\-trol Po\-li\-cy}\xspace}

\newcommand{\ContextDataCollector}{\textsf{Da\-ta\ Col\-lec\-tor}\xspace}
\newcommand{\ContextClassifier}{\textsf{Clas\-si\-fi\-er}\xspace}
\newcommand{\ContextProfiler}{\textsf{Pro\-filer}\xspace}

\newcommand{\flaskdroid}{\emph{FlaskDroid}\xspace}
\newcommand{\userspaceobjectmanagers}{\textsf{User\-Space\ Ob\-ject\ Ma\-na\-gers}\xspace}

\newcommand{\usom}{\textsf{USOM}\xspace}
\newcommand{\usoms}{\textsf{USOMs}\xspace}
\newcommand{\ContextProviders}{\textsf{ContextProviders}\xspace}
\newcommand{\ContextProvider}{\textsf{ContextProvider}\xspace}
\newcommand{\CameraService}{\textsf{CameraService}\xspace}
\newcommand{\SensorService}{\textsf{SensorService}\xspace}
\newcommand{\SystemServer}{\textsf{SystemServer}\xspace}
\newcommand{\flaskdroids}{\emph{Flask\-Droid's}\xspace}

\newcommand{\untrusted}{\textit{un\-trus\-ted}\xspace}
\newcommand{\trusted}{\textit{trus\-ted}\xspace}

\newcommand{\PosObs}{\mathit{pos}}

\newcommand{\StayPoint}{\mathit{sp}}
\newcommand{\StayPointRadius}{r_{\StayPoint}}
\newcommand{\StayPointMinDuration}{\mathit{t\_min_{\StayPoint}}}
\newcommand{\StayPointMaxGap}{\mathit{t\_gap_{\StayPoint}}}

\newcommand{\StayPointAvgPos}{\mathit{pos_{\bar{\StayPoint}}}}

\newcommand{\GPSCoIMaxSize}{\mathit{gps_{max}}}
\newcommand{\CoIMinFreq}{\mathit{f\_min_{coi}}}
\newcommand{\CoIMinTime}{\mathit{t\_min_{coi}}}
\newcommand{\CoI}{\mathit{C}}
\newcommand{\AllCoIs}{\mathcal C}

\newcommand{\Duration}{\textsf{dur}}

\newcommand{\WiFiObs}{\mathit{rf}}
\newcommand{\WiFiSnapshot}{\mathit{wifi}}
\newcommand{\WiFiSnapshotMaxTime}{\mathit{t\_max_{wifi}}}
\newcommand{\JaccardDistance}{\mathit{J_\delta}}
\newcommand{\WiFiStayPoint}{\mathit{wifi\_sp}}
\newcommand{\CharacteristicSet}{\textsf{char}}

\newcommand{\Device}{\mathit{d}}
\newcommand{\AllDevices}{\mathcal D}
\newcommand{\DeviceObs}{\mathit{bt}}

\newcommand{\Visit}{V}
\newcommand{\VisitMaxGap}{\mathit{\epsilon_{\Visit}}}
\newcommand{\AllVisits}{\mathcal{\Visit}}

\newcommand{\Encounter}{E}
\newcommand{\EncounterMaxGap}{\epsilon_{\Encounter}}
\newcommand{\AllEncounters}{\mathcal{\Encounter}}

\newcommand{\DeviceContext}{\mathit{D}}
\newcommand{\LocationContext}{\mathit{L}}

\newcommand{\FamiliarDevices}{\mathcal{D}_{fam}}
\newcommand{\FamDeviceMinTime}{\mathit{t\_min_{famdev}}}
\newcommand{\FamDeviceMinFreq}{\mathit{f\_min_{famdev}}}

\newcommand{\CoIProfile}{\mathit{CoIs}}
\newcommand{\DeviceProfile}{\mathit{Devs}}
\newcommand{\VisitFeatures}{\mathit{P}}
\newcommand{\DeviceFeatures}{\mathit{O}}

\newcommand{\CoIVisitCount}{\mathit{visits}}
\newcommand{\CoIVisitTime}{\mathit{dur}}
\newcommand{\DeviceEncounterCount}{\mathit{enc}}
\newcommand{\DeviceEncounterTime}{\mathit{dur}}

\newcommand{\ContextFeature}{f}

\newcommand{\MaxVisitTimeGPSCoI}{\ContextFeature_{\text{max}_\text{dur}}^\text{GPS}}

\newcommand{\MaxVisitTimeWiFiCoI}{\ContextFeature_{\text{max}_\text{dur}}^\text{WiFi}}
\newcommand{\MaxVisitWiFiCoI}{\ContextFeature_{\text{max}_\text{freq}}^\text{WiFi}}
\newcommand{\DevicesInContext}{\ContextFeature_{\text{num}}^\text{BT}}
\newcommand{\FamDevicesInContext}{\ContextFeature_{\text{fam}}^\text{BT}}
\newcommand{\AvgEncTimeFamDevices}{\ContextFeature_{\text{fam}_\text{avg-time}}^\text{BT}}
\newcommand{\AvgEncFamDevices}{\ContextFeature_{\text{fam}_\text{avg-freq}}^\text{BT}}

\newcommand{\thirdparty}{$3^\text{rd}$-party\xspace}

\definecolor{lightgray}{gray}{0.9}
\definecolor{gray}{rgb}{0.5, 0.5, 0.5}
\definecolor{personcolor}{rgb}{0.9, 0.7, 0.7}

\makeatletter\newenvironment{graybox}{%
\begin{lrbox}{\@tempboxa}\begin{minipage}{\columnwidth}}{\end{minipage}\end{lrbox}%
\colorbox{lightgray}{\usebox{\@tempboxa}}
}\makeatother



\newenvironment{attention}{%
\begin{tabular}{m{4mm}|m{2mm}m{115mm}}
\textbf{\rotatebox{90}{\mbox{Note}}} & &
\begin{minipage}[t]{\linewidth}
}{
\end{minipage}
\end{tabular}
}


%% file: intro.tex
Mobile devices today are equipped with a wide variety of sensors for sensing the context of the device.
Applications that make use of this information are becoming increasingly
popular. Examples include location-based applications like \emph{Foursquare}
and \emph{Tencent WeChat}, augmented reality applications like \emph{Layar},
\emph{Wikitude}, \emph{Google Goggles} and \emph{HERE City Lens} among many more.  Even
mainstream applications like social network apps support context-based
enhancements.

The improved sensing capabilities of modern
smart devices also provide an attractive attack surface against the contextual privacy of users, as the
recent development of sensory malware shows: malicious code, typically a Trojan Horse appearing to be a legitimate app,  uses the
sensors of the device to extract sensitive information from the
surroundings of the user. 
Context-aware access control can be used to encounter this threat by limiting
the access of untrusted \thirdparty applications to context information.

Various context-aware access control mechanisms and systems have been proposed. Some of these works are based on modifications of the RBAC model~\cite{Covington2002, Damiani2007} in which context-awareness is realized through roles that are triggered based on context parameters. Other approaches use explicit policies conditioned on contextual parameters~\cite{Sadeh2009, CONUCON, Conti2012, Bell2012} or rules for expressing higher-level contextual preferences~\cite{Hull2004}. 

All of these works are based on \emph{user-de\-fined} or \emph{pre-defined} policies. User-defined policies are very laborious to set up and maintain. This can be encountered with pre-defined policies provided by system administrators or app vendors, but these are inaccurate and inflexible: they cannot adequately capture and adapt to the highly personal and dynamic nature of individual users' contexts and privacy preferences. In our work, we overcome these deficiencies by using automatic context classification as a basis for access control decisions instead of static access control policies.

Furthermore, improved sensing capabilities also provide better possibilities to encounter threats arising from the context, like the threat of physical device misuse.
Current static device locking methods severely deteriorate device usability with unnecessary password prompts in low-risk contexts, causing many users to leave their device unprotected~\cite{Siciliano2013}. Some earlier works attempted to improve device locking 
by using context information to probabilistically determine a level of confidence in the user's authenticity~\cite{riva2012progressive, Hayashi:CASA}. We take a different approach: 
we do not try to authenticate users, but adjust device locking criteria according to the perceived security risk level of a context.


To ground our work we use an interdisciplinary approach in which a sociological study is used to identify the concerns and perceptions of smartphone users in different contexts.

\textbf{Contributions.} Our contributions  are as follows:

\begin{itemize}
\item We introduce \ouremphname, the first
  context-aware access control framework for smartphones
  that uses \textbf{context profiling} and \textbf{automatic, adaptive and personalized context classification}
  for making con\-text-aware access control decisions.

\item We apply our framework to two important smartphone use cases:
  \textbf{protecting against device misuse}, and \textbf{defending against sensory malware}.
  \ourname, however, is also applicable to a wide range of other security and
  privacy-related use cases.

\item We apply \textbf{context profiling} and \textbf{machine learning} techniques on real-world data collected in a user study and evaluate the efficiency of automatic context classification in addressing the aforementioned threats.

\item We integrate \ourname with the \flaskdroid~\cite{flaskdroidUsenix} architecture for fine-grained access control enforcement on the Android platform in order to realize the first adaptive, personalized and context-aware access control system of its kind for mobile devices.

\end{itemize}
The rest of this paper is structured as follows. A description of the problem and the \ourname framework is given in Sections \ref{problem} and \ref{framework}. In Section \ref{userstudy} we introduce the results of a sociological study on users' concerns and perceptions related to smartphones and present use cases addressing these concerns in Section \ref{usecases}. The design and implementation of our context model are described in Sections \ref{contextmodel} and \ref{implementation}. The results of a user study evaluating the performance of context classification are presented in Section \ref{evaluation}, and Section \ref{accesscontrollayer} shows its integration with the \flaskdroid architecture. 
After summarizing related work in Section \ref{relatedwork}, we conclude with outlines of future work in Section \ref{conclusions}.

%% file: problem.tex
While the idea of context-aware access control is not new \cite{Covington2002,Hull2004,Damiani2007,Conti2012}, currently proposed solutions mostly rely on policies specifying access control rules conditioned on values of contextual parameters.

User-specified policies have the potential to correctly reflect the user's true security and privacy preferences, but
the amount of work required to set up and maintain a comprehensive set of context-dependent policies is high. Average users of mobile devices are hardly willing to spend significant amounts of time maintaining their policy set. In addition, 
it is questionable, whether regular users are capable to fully understand the implications of the policy settings they define. 
A study concerning location sharing policies~\cite{Sadeh2009} showed that the initial accuracy of location disclosure rules specified by users was only $59\%$ and improved to $65\%$ after users modified the rules based on a review of concrete enforcement decisions resulting from these rules.
A recent study on the users' willingness to share their data with prominent single-sign on services \cite{Bauer2013} showed that the \emph{majority} of users  did not correctly understand the sharing implications of a sign-on dialog directly presented to them. These results suggest that the users' practical ability to control their security and privacy settings is limited.


An approach to tackle this problem of complexity and required user effort is to simplify things for users by pre-defined policies provided by administrators or app vendors. While this effectively reduces the user's burden, it fails to address the individual needs and privacy preferences of users. Pre-defined policies are by necessity only generalizations and not capable of accurately capturing concrete contexts and situations in individual users' lives. Also the fact that the privacy preferences of users may vary significantly from person to person\citation{citationNeeded} can be very difficult to address using pre-defined policies.


It seems also unlikely that the privacy and security implications of pre-defined policies would be fully understood by regular users.
 This may lead to undesired situations: users under- or overestimating the level of privacy and security protection that the pre-defined policy provides to them. 
User- or pre-defined policies alone are not sufficient to capture the highly dynamic and highly personal nature of a user's context (i.e., the ambient environment the device finds itself in) due to the problems outlined above. We aim at encountering these deficiencies by designing a context-aware access control framework that 
captures and adapts to the user's perception of the context and performs automatic context classification for making fine-grained context-aware access control decisions without the need for users to explicitly define contextual constraints.
In the following section we describe the \ourname framework in more detail.

%% file: framework.tex
\begin{figure}
\includegraphics[width=\columnwidth]{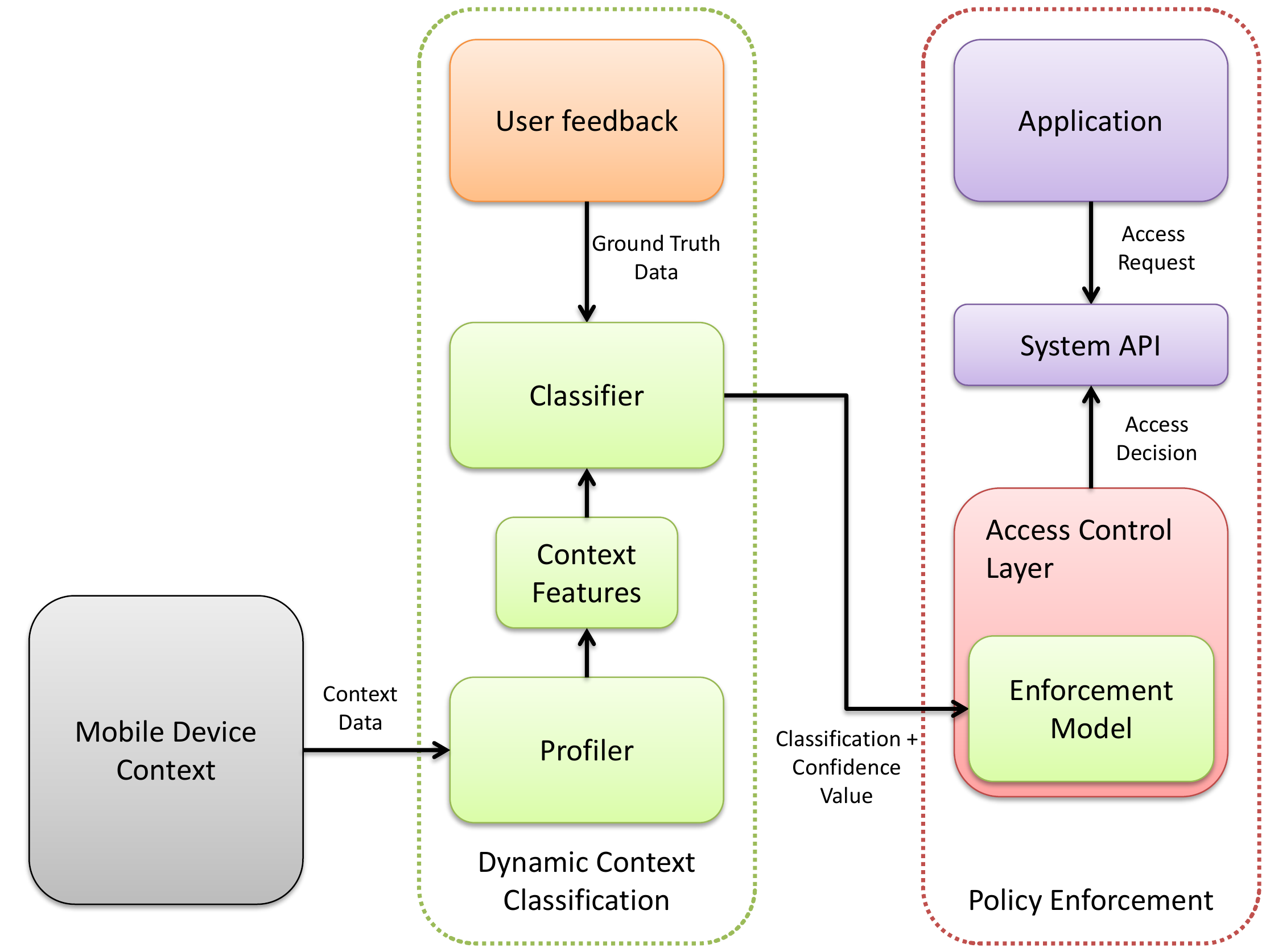}
\caption{Context-based access control enforcement in \ourname}
\label{fig:framework}
\end{figure}

The \ourname framework provides context-aware access control decisions by performing automatic classification of the context with regard to its security-relevant properties. The classification is based on machine learning models and user feedback providing ground truth information for training these models.
Figure \ref{fig:framework} shows a high-level overview of the framework.
%

The framework architecture is driven by context data observed with the sensors of the mobile device. The data are fed to a \ContextProfiler to calculate features describing the context. The \ContextProfiler implements a context model and aggregates profiles for relevant objects (e.g., significant places of the user or devices the user encounters) in the model. The \ContextProfiler evaluates incoming observations based on the profiles and the context model, and periodically calculates feature vectors characterizing the current context of the user. 

Apart from sensed context data, the framework obtains input through user feedback. This feedback can be derived from explicit user interaction (e.g., feedback given through the device's UI) or by monitoring the user's actions.

The \ContextClassifier uses the context feature vectors and user feedback events to train and update the context classification models. Once the models have been trained, they are used to classify new observations with regard to the context's security and privacy-relevant properties. 
The classification estimate of the \ContextClassifier and its associated confidence value are forwarded to the \accesscontrollayer, which takes them into account when making access control decisions. 
The framework can be used to address any use case for which contextual factors play a role, and it can accommodate any sensors, resources, functionalities, communication links or other data objects for which the \accesscontrollayer provides enforcement support.

In the following, we instantiate the \ourname framework for protecting the privacy of the user. We consider
two central use cases we identified and demonstrate the applicability and effectiveness of our approach by evaluating it with real-world contextual data obtained from a user study.

%% file: userstudy.tex
To investigate smartphone users' perceptions and concerns with regard to their smartphones in different contexts, we conducted a user survey following a \emph{Mixed Methods}~\cite{tashakkori2003handbook} approach commonly applied in sociological studies. We designed a sta\-tis\-ti\-cal-quantitative survey \cite{QuantitativeTradition} using quantitative questions to identify facts by statistical analysis~\cite{alan2007business} combined with open-ended, qualitative questions for investigating the underlying reasons for the users' perceptions.

The survey was answered by 122 participants aged 18-56 including people from different household types and representing different organizational positions. The participants were recruited using word-of-mouth, electronic communications like e-mail and social networks, targeting particularly smartphone users, to obtain a cross-sectional sample of different age groups and backgrounds. Thus, the set of respondents in our sample is representative of the target group of our framework, namely
active smartphone users.

The survey contained questions on which contexts and contextual factors users deem relevant for their perceptions of  contextual privacy and security. The answers to the open-ended questions confirmed our initial assumptions that two major context-related concerns dominate the users' minds. Firstly, people are concerned about \emph{device misuse}, i.e., that their device is stolen and/or misused without their knowledge. The second concern relates to \emph{privacy exposure}: users fear that private or confidential context information related to their life is revealed to unauthorized parties.

In some earlier studies on contextual behavior patterns of mobile device users, two central contexts have been used: ``home'' and ``work''~\cite{Verkasalo_contextual_2008}. Also lifetime studies in social sciences suggest that these contexts are the most important contexts in average users' lives~\cite{Sennett1992uses}. Therefore, we included dedicated questions about these context types as prototypes of very familiar contexts.
Tables \ref{tab:homework} 
and \ref{tab:people-in-context} show the results.\\

\subsection{Survey Results}
\noindent
\textbf{Risk of device misuse.}
As can be seen in Table \ref{tab:homework}, the majority of people ($94\%$ and $55\%$, respectively) perceive ``home'' and ``work''  as having a low risk of misuse. This is explained by answers to open-ended questions, like ``($\ldots$) places like home or office tend to remain secure regardless of time of day'', supporting the intuition that familiar places are commonly perceived as having a low risk of misuse.
While it is not surprising that ``home'' is perceived as safe, perceptions of ``work'' are more diverse~\cite{cohen1992escape}, as reflected in our survey results. A significant fraction of people 
perceive ``work'' as a context with high risk of misuse. The reasons for this are indicated in several answers to the open-ended questions, e.g., ``At work there are people around that I don't know and I don't trust them.'' This suggests that the familiarity of the context alone is not a sufficient indicator for estimating the risk level, also the people in context play an important role.

From Table \ref{tab:people-in-context} we can see that the people present in the context clearly affect the perceived risk of device misuse in that context. A clear majority of respondents 
stated that a low or high risk of device misuse is dependent on the persons present. From the responses to open-ended questions, we identified that the feeling of low risk is particularly caused by people that are \emph{familiar} to the user (e.g., ``I trust my friends and colleagues''). Similarly, the presence of \emph{unfamiliar} persons in the context was indicated as a reason for perceiving the context as having a high risk of device misuse (e.g., ``Where there are people around me that I don't know, I don't feel secure'', or ``Unknown people represent threat'').

This underlines the fact that location information alone is not a sufficient basis for access control decisions. Also other context information affecting the perceptions of security need to be taken into account.\\
\noindent
\textbf{Privacy exposure.}
The data in Table \ref{tab:homework} suggest that a significant fraction of respondents feel 
that a familiar context also has high privacy exposure, i.e., contains private or confidential  context information related to the user.

Table \ref{tab:people-in-context} shows, however, that the perception of privacy exposure does not appear to be affected by the presence of people. More respondents 
believe that the people present in the context do not contribute to the level of privacy exposure of the context.
\begin{table}[]
  \centering
  \caption{Perceptions of Home vs. Work\label{tab:homework}}
  {\small
    
    \begin{tabular}{||l||c|c||c|c||}
      \hline
&\multicolumn{2}{c||}{Privacy exposure} & \multicolumn{2}{c||}{Risk of misuse}\\
      Context & high & low & low & high  \\
      \hline
     Home & 46\% &  17\% & 94\%& 4\% \\
     Work  & 42\% & 21\% & 55\% & 40\% \\ 
      \hline
    \end{tabular}
  }
\vspace{0.3cm}
\caption{Influence of people on the perceived privacy exposure and risk of misuse in the context}
\label{tab:people-in-context}
{\small
    
    \begin{tabular}{||l|c|c||c|c||}
   \hline
      Question  & Yes & No \\
      \hline
     Low risk of misuse depends on people & 66\% &  14\% \\
     High risk of misuse depends on people  & 68\% & 11\% \\ 
     High privacy exposure depends on people & 39\% & 43\% \\
     Low privacy exposure depends on people & 36\% & 42\% \\
      \hline
    \end{tabular}
  }
  
\end{table}

\noindent
\textbf{Conclusions and Discussion.}
Based on the analysis of the survey results above, we 
see that two main factors affect users' perceptions on privacy exposure and risk of misuse in contexts: the familiarity of the context itself and the familiarity of persons present in the context.
%
Therefore, we need to design our context model in a way that we can (1)~identify relevant contexts and model their familiarity, and, (2) track encounters with other persons and measure their familiarity by observing their mobile devices. In Section \ref{contextmodel} we construct such a context model.



The analysis presented above supports our \emph{common} understanding and shows which factors typically influence the perception of contexts.
However, exceptions exist, like, e.g., those $4\%$ who consider home 
a high-risk environment. Reasons for this can be various. Answers to open-ended questions suggest that even in familiar contexts the perception of risk of misuse changes when untrusted people appear, or, it is caused by what we call the ``Toddler Scenario'', i.e.,  the influence of familiar people (e.g., a young child, or spouse) considered as ``clueless'' or ``honest but curious'', causing a person to consider the risk of device misuse significant also in familiar surroundings.
To investigate these special cases, more detailed questions on familiar contexts would need to be included. However, sociological literature suggests that questions about such contexts are perceived as intrusive and will therefore not be answered \cite{beck1992risk,cohen1992escape}. This concern was also reflected in some of the feedback we received about the online questionnaire. Hence, and because exceptional cases seem to be a marginal phenomenon, we focus on the common cases and address more exceptional cases in subsequent studies.

%% file: usecases.tex
%
Based on the user survey, 
we focus on the most prominent concerns expressed by the users: fear of device misuse and disclosure of private or confidential context information. To mitigate these concerns, we identified following use cases.

\subsection{Misuse Protection: Usable Device Lock}
Several surveys~\cite{Camp2012,Siciliano2013}
point out that many mobile users do not use device locks (also known as idle screen locks) to
protect their phones even though that would effectively protect against device misuse.  One reason for this is that screen locks and other
similar access control techniques on mobile devices today
are both too inflexible and hard to
use. 
A solution could be to make the locking mechanism more usable, so that users would be more willing to use device locking. The approach taken by Gupta et al.~\cite{DBLP:conf/socialcom/0002MAN12} was to use context data to adapt the locking time out of the device lock in different contexts. We adopt this approach and want to use the estimated \emph{risk of device misuse} in a context as a means to decide, whether and how fast to lock the device in case it is not used.
\\
\noindent 
\textbf{Adversary model.}
 For this use case, the adversary is a person in the context with physical access to the device.  The person may be malicious (a thief) or honest-but-curious (a colleague or sibling) or ``clueless" (a small child).\\
\noindent 
\textbf{Goal.} We want to protect the applications and data on the device
from potential threats in the context by limiting the potential damage
arising from someone physically accessing the device without the
user's approval.  
Therefore, we want to minimize the chances that an unauthorized person in the context has access to the user's data.  We do this by configuring the device lock dynamically based on the risk of device misuse in the context, while trying to strike a balance between maximizing protection on one hand and minimizing
user annoyance of having to unlock the device in low-risk contexts on the other hand.

\subsection{Resisting Sensory Malware}
Sensory malware is an emerging class of malicious applications (typically Trojans) that use the context sensors of a mobile device to collect potentially sensitive information from the user's context. Prominent examples of sensory malware are \emph{Stealthy Video Capturer}~\cite{Xu2009} (video via camera),
\emph{(sp)iPhone}~\cite{Marquardt2011} (keystrokes via accelerometer) \emph{Soundcomber}~\cite{Schlegel2011}
(spoken secrets via microphone), or the recent \emph{PlaceRaider}~\cite{Templeman2013} Trojan
(3D models via camera).  Users may also have granted sensor
access privileges to benign apps which use them too
intrusively: for instance 
an augmented reality app 
may take
pictures of surroundings even when the user is not actively using
augmented reality, as a means of enriching the app vendor's data
collection. \\
\noindent\textbf{Adversary Model.} For this use case, the adversary is an app already installed on the device.  We assume that the application has been granted the necessary privileges during installation and has therefore access to the contextual sensors on the mobile device, but cannot circumvent the access control system\footnote{Malware that uses operating system (root) exploits to circumvent the enforcement of the context-aware access control system is outside the scope of this paper.}.  The application may be a Trojan Horse (e.g., sensory malware) or a benign but somewhat intrusive application. \\
\noindent\textbf{Goal.} We aim at protecting sensitive information in the context of the device from the adversary.  We do this by preventing or limiting the ability of the adversary to gather information from contexts with high \emph{privacy exposure}, i.e., contexts that contain information that the user would want to protect from the adversary. Such information can be either \emph{private}, i.e., information about the user herself, or, \emph{confidential}, i.e., other sensitive information not directly related to the user.
The user's home (private) and workplace (confidential) are examples of typical contexts with high privacy exposure.

%% file: contextmodel.tex
In this section, we present a context model used to extract context features reflecting the familiarity of contexts and the persons in the context. 
The context features are input for the \ContextClassifier and used for classifying contexts as having high or low privacy exposure and/or risk of misuse. 
The context model is based on two main concepts: \emph{Contexts of Interest (CoI)} for modelling the familiarity of contexts and \emph{Bluetooth devices} for modelling familiar and unfamiliar people in context.

\subsection{Detection of Contexts of Interest (CoIs)}
\label{coi_detection}
For our purpose, \emph{Contexts-of-Interest} (CoIs) correspond to locations that a user often visits  and/or spends a significant amount of time in, e.g., home, workplace, grocery store, etc.
We consider two kinds of CoIs: \emph{GPS-based CoIs} which are geographical areas on the surface of the earth, and \emph{WiFi-based} CoIs that are characteristic sets of WiFi access points usually observed in a specific place and thus identifying the RF environment there. GPS CoIs capture significant places of the user in outdoor areas, and WiFi CoIs cover also indoor locations in urban areas, where GPS can't be used but coverage of WiFi access points typically is available. By combining both types of CoIs, we can identify and detect most significant places that users typically visit.
\subsubsection{GPS-based CoIs}
To identify GPS-based CoIs, we adopt the notions of \emph{stay points} and \emph{stay regions} as introduced by Zheng et al.~\cite{Zheng2010} and developed further by Montoliu et al.~\cite{Montoliu2013}.
The identification of GPS-based CoIs is based on position observations
obtained via GPS. 

The sequence of GPS observations is divided into \emph{GPS stay points}, which represent visits of the user to different places, during which the user stays within a radius of $\StayPointRadius = 100$ m from the first GPS observation. In order for a visit to be considered a stay point, the visit is also required to last longer than $\StayPointMinDuration = 10$ min and not to contain observation gaps longer than $\StayPointMaxGap = 5$ min.

We calculate for each stay point an average position $\StayPointAvgPos$ as the average of all position observations belonging to the stay point, i.e., $\StayPointAvgPos = (lat_{\bar{sp}}, lon_{\bar{sp}})$, s.t. $lat_{\bar{sp}} = \frac{\sum_{k = 1}^{n}{lat_k}}{n}$, and $lon_{\bar{sp}} = \frac{\sum_{k = 1}^{n}{lon_k}}{n}$. The average position of a stay point represents the predominant location where the user has been located during her visit to the stay point. 

The average positions $\StayPointAvgPos$ of individual stay points are aggregated to form rectangular geographical areas of at most $\GPSCoIMaxSize = 100$ m width and length. An area is a \emph{GPS-based Context-of-Interest}, if (i) the user has visited the area more than $\CoIMinFreq = 5$ times and (ii) has spent  longer than $\CoIMinTime = 30$ min in total in the area. 

\noindent\textbf{Example.} As an illustrative running example, let us consider a user who regularly commutes between her workplace and home.
Other places she regularly visits are a grocery store and a public sports facility. 
 She usually carries her smartphone with her, which continuously senses her GPS location and other context data.

When the user goes to the grocery store and stays there for 15 minutes, i.e., longer than $\StayPointMinDuration = 10$ min and moves only within a radius of $\StayPointRadius = 100$ m, a stay point $\StayPoint$ of duration $\Duration(\StayPoint) = 15$ min will be generated. The average of all position observations $\PosObs_i$ during the stay point visit will be the stay point average position $\StayPointAvgPos$, most likely located in or near the grocery store.
Waypoints along her daily commuting routes, however, would not generate any stay points, since on her way she does not spend sufficiently long time in the same limited area.

If our user visits the grocery store 10 times and stays each time for 15 minutes, ten stay points will be generated.
These average positions will be aggregated into a GPS-based CoI $\CoI$, because their total stay duration of 2 hours and 30 minutes is longer than the required $\CoIMinTime = 30$ min and there are more than the required $\CoIMinFreq = 5$ stay points falling inside the CoI. The area of the CoI will be the smallest rectangle containing all the stay point average positions $\StayPointAvgPos$. 



\subsubsection{WiFi-based CoIs}
For identifying WiFi-based CoIs, WiFi access point observations $\WiFiObs_i$ are used. Each observation consists of the MAC address of a detected WiFi access point and the timestamp of the observation. 
The sequence of individual WiFi observations is divided into \emph{WiFi snapshots}, which are subsequences corresponding to observations obtained during a single WiFi scan of duration $\WiFiSnapshotMaxTime = 10$ sec.

Following the notion of stay points for GPS observations, we extend this concept to WiFi and divide the sequence of WiFi snapshots into so-called \emph{WiFi stay points}.
The similarity between snapshots is determined by calculating the \emph{Jaccard distance}\footnote{The Jaccard distance measures the dissimilarity between two sets $A$ and $B$ as $\JaccardDistance(A, B) = \frac{\vert A \cup B \vert - \vert A \cap B \vert}{\vert A \cup B\vert}$}
between the first snapshot and subsequent snapshots one-by-one. As long as the Jaccard distance between the snapshots is less than or equal to $0.5$, which means that the intersection of the snapshots is at least as large as half of their union, the subsequent snapshots are assigned to the stay point. The staypoint is considered complete, if the Jaccard distance to new WiFi snapshots grows beyond $0.5$ or there is a gap between consecutive WiFi snapshots that is longer than $\StayPointMaxGap$.
These criteria for WiFi stay points were selected, because it is not uncommon that WiFi access points are missed in scans~\cite{Dousse2012}. This is apparently not dependent on the signal strength of the missed access point, so one needs to take into account that even very strong access point beacons will be missed from time to time.

A WiFi stay point has a characteristic set of access points $\CharacteristicSet(\WiFiStayPoint)$ that includes those access points that occur at least in half of all WiFi snapshots of the stay point. 
A set of access points  is a \emph{WiFi-based CoI}, if there are at least $\CoIMinFreq$ WiFi stay points having this set of access points as their characteristic set of access points, and the stay points have a duration of at least $\CoIMinTime$ in total.

\noindent
\textbf{Example.}
When our example user arrives at her workplace, a WiFi snapshot $\WiFiSnapshot$ is recorded. This snapshot and following snapshots having a Jaccard distance of less than or equal to $0.5$ to the first one form a WiFi stay point $\WiFiStayPoint$, given that the time difference of the first and last snapshot is greater than $\StayPointMinDuration$ and there are no gaps in the WiFi snapshot observations longer than $\StayPointMaxGap$. The characteristic set $\CharacteristicSet(\WiFiStayPoint)$ of access points of this stay point consists of access points mostly observed at the workplace.
During subsequent visits to the workplace, more WiFi stay points with the same characteristic set will be generated. If at least $\CoIMinFreq$ such stay points have been observed and the total visit duration $\Duration(\WiFiStayPoint)$ of these stay points reaches $\CoIMinTime$, the characteristic set constitutes a WiFi-based CoI $\CoI$ for the user's workplace. 


\subsection{Context Detection}
Once the GPS- and WiFi-based CoIs have been identified, new incoming GPS, WiFi and Bluetooth observations can be used to identify the location context and social context of the user at any point in time.
\subsubsection{Location context}
The location context of the user is defined in terms of the CoIs that the user visits at a specific point in time.
\begin{definition}[Visits]
A user's \emph{visit} $\Visit_{\CoI}$ to a GPS-based CoI $\CoI = (lat_{min}, lon_{min}, lat_{max}, lon_{max})$ is a sequence of position observations $\PosObs_i = (lat_i, lon_i)$ falling within the CoI and having timestamps at most $\VisitMaxGap$ apart from each other:
$\Visit_{\CoI} = (\PosObs_1, \PosObs_2, \ldots, \PosObs_n)$, where 
$\forall \PosObs_i \in \Visit_{\CoI}: lat_{min} < lat_i \wedge lon_{min} < lon_i  \wedge lat_i < lat_{max} \wedge lon_i < lon_{max}$, and 
$\forall i, 1 < i \leq n: t(\PosObs_i) - t(\PosObs_{i-1}) < \VisitMaxGap$. 
Similarly, a visit $\Visit_{\CoI}$ to a WiFi-based CoI $\CoI$ is a sequence of WiFi snapshots $\WiFiSnapshot$ falling within the CoI and having timestamps at most $\VisitMaxGap$ apart from each other. That is, 
$\Visit_{\CoI} = (\WiFiSnapshot_1, \WiFiSnapshot_2, \ldots, \WiFiSnapshot_n)$, where $\JaccardDistance(\CoI, \WiFiSnapshot_i) \leq 0.5$ and $\forall i, 1<i \leq n: t(\WiFiSnapshot_i) - t(\WiFiSnapshot_{i-1}) < \VisitMaxGap$. We denote the set of all visits $\Visit_{\CoI}$ of the user to CoI $\CoI$ with $\AllVisits_{\CoI}$.
\end{definition}

\begin{definition}[Location Context]
A \emph{location context} $\LocationContext_t$ at timestamp $t$ is the set of CoIs $\CoI$ that the user is visiting during that point of time.
\end{definition}
Note, that CoIs can be overlapping, which means that a user can be visiting several CoIs simultaneously. If the user is not visiting any of the CoIs at a specific point in time, the corresponding location context will be empty.
%

\subsubsection{Social context}
In order to capture \emph{people} in the user's surroundings, we observe their mobile devices that can be sensed through proximity sensing technologies like Bluetooth (BT). Bluetooth has a range of approximately 30 meters given a direct line of sight, so its physical properties quite well reflect our notion of a context comprising the space immediately observable by the user (e.g., a room). Bluetooth has been commonly used in ubiquitous computing literature to model the presence of persons in a perimeter (cf., e.g., \cite{Dousse2011}). To capture only devices that are typically carried by persons, we filter the BT observations by their device class so that we consider only mobile devices like cell phones, headsets, PDAs and other portable devices.

Some users may not keep Bluetooth enabled
and discoverable on their devices or always carry their devices
with them. Therefore, we will not always be able to reliably
detect the presence of all persons in the context using Bluetooth alone.
However, this is not necessary, since our probabilistic
framework utilizes Bluetooth as one factor for identifying
the type of context the user is in and not as a 'tripwire' for
detecting potentially malicious users. Especially in public
contexts where many persons are present and the likelihood
that at least some Bluetooth devices can be detected is high,
Bluetooth works well as a context classification factor. In
addition, familiar, known devices (e.g. devices of family and
friends) can be polled even if they are in hidden mode, if the
BDADDR of the target device is known.  For example, two devices
that are paired can detect each other this way even if they
remain invisible to other devices.


The social context is defined in terms of the devices that are detected in the user's context at a specific point in time.
\begin{definition}[Encounters]
An \emph{encounter} $\Encounter_\Device$ of a user with a device $\Device$ is a sequence of Bluetooth observations $\DeviceObs_i$ of device $\Device$ with timestamps that are at most $\EncounterMaxGap$ apart from each other:
$\Encounter = (\DeviceObs_1, \DeviceObs_2, \ldots, \DeviceObs_n)$, where $\forall i, 1< i \leq n: \DeviceObs_i = \Device \wedge t(\DeviceObs_i) - t(\DeviceObs_{i-1}) < \EncounterMaxGap$. We denote the set of all encounters of the user with a device $\Device$ with $\AllEncounters_{\Device}$.
\end{definition}

When our example user arrives at her workplace, her device obtains a Bluetooth observation $\DeviceObs_1 = \Device$ of her colleague's device $\Device$. This observation and any subsequent device observations $\DeviceObs_i = \Device$ of the colleague's device form
an encounter $\Encounter_{\Device}$ with the colleague's device, as long as the time distance between consecutive device observations is less than $\EncounterMaxGap = 5$ minutes.
The purpose of allowing gaps of this size is to be able to handle missed device observations not uncommon with Bluetooth sensing.

%

\begin{definition}[Device Context]
The \emph{device context} $\DeviceContext_t$ at timestamp $t$ is the set of devices $\Device$ that are encountered during that point of time.
\end{definition}
\begin{definition}[Familiar Devices]
\label{famdevs}
The set of \emph{familiar devices} $\FamiliarDevices$ is the set of all such devices that the user has encountered at least $\FamDeviceMinFreq$ times and for which the total duration of the encounters is at least $\FamDeviceMinTime$. 
\end{definition}

Familiar devices $\Device$ for our example user would be the mobile devices of familiar people like her spouse or her colleagues at work which she has encountered more often than $\FamDeviceMinFreq = 5$ times and the total duration of these encounters is longer than $\FamDeviceMinTime = 30$ minutes.



\subsection{Context Profiles}
\label{contextprofiles}
Based on the above context model, context profiles are aggregated for the user: a \emph{CoI profile} $\CoIProfile$ and a \emph{device profile} $\DeviceProfile$. The CoI profile $\CoIProfile = \{\AllCoIs, \VisitFeatures\}$ consists of the set of all identified CoIs $\AllCoIs$, 
and a mapping $\VisitFeatures: \AllCoIs \to \mathbb N \times \mathbb R, \CoI \mapsto (\CoIVisitCount_\CoI, \CoIVisitTime_\CoI)$ providing the total amount $\CoIVisitCount_\CoI$ and total duration $\CoIVisitTime_\CoI$ of visits to each CoI $\CoI \in \AllCoIs$.

Similarly, the device profile $\DeviceProfile = \{ \AllDevices, \FamiliarDevices, \DeviceFeatures \}$ consists of the set of all encountered devices $\AllDevices$, the set of familiar devices $\FamiliarDevices$ and a mapping $\DeviceFeatures: \AllDevices \to \mathbb N \times \mathbb R, \Device \mapsto (\DeviceEncounterCount_\Device, \DeviceEncounterTime_\Device)$ providing the total amount $\DeviceEncounterCount_\Device$ and total duration $\DeviceEncounterTime_\Device$ of encounters with each device $\Device \in \AllDevices$.

\subsection{Context Features}
Based on the context model, we define following features:
%

\noindent
\textbf{Context familiarity features.} \\
Let $\AllCoIs^{\text{GPS}}$ denote the subset of all GPS-based CoIs and $\AllCoIs^{\text{WiFi}}$ the subset of all Wifi-based CoIs in $\AllCoIs$. Then we have:\\ 
$\MaxVisitTimeGPSCoI$: maximum total visit time of any GPS-based CoI in current location context
\begin{displaymath}
\MaxVisitTimeGPSCoI(t) = \left\{ \begin{array}{rl}
\max_{\CoI \in \{ \AllCoIs^{\text{GPS}} \cap \LocationContext_t \} }{\CoIVisitTime_\CoI}, & \LocationContext_t \cap \AllCoIs^{\text{GPS}} \neq \emptyset \\
0, & \text{otherwise}\\
\end{array} \right .
\end{displaymath}
$\MaxVisitTimeGPSCoI$: number of visits to the GPS-based CoI with the maximum total visit time\\
\begin{displaymath}
\MaxVisitTimeGPSCoI(t) = \left\{ \begin{array}{rl}
\CoIVisitCount_{\CoI_i}, & i = \arg \max_{\CoI \in \{ \AllCoIs^{\text{GPS}} \cap \LocationContext_t \} }{\CoIVisitTime_\CoI} \wedge \\ & \LocationContext_t \cap \AllCoIs^{\text{GPS}} \neq \emptyset\\
0, & \text{otherwise}\\
\end{array} \right .
\end{displaymath}
$\MaxVisitTimeWiFiCoI$: maximum visit time of any WiFi-based CoI in the location context\\
\begin{displaymath}
\MaxVisitTimeWiFiCoI(t) = \left\{ \begin{array}{rl}
\max_{\CoI \in \{ \AllCoIs^{\text{WiFi}} \cap \LocationContext_t \} }{\CoIVisitTime_\CoI}, & \LocationContext_t \cap \AllCoIs^{\text{WiFi}} \neq \emptyset\\
0, & \text{otherwise}
\end{array} \right .
\end{displaymath}
$\MaxVisitWiFiCoI$: number of visits to the WiFi-based CoI with the maximum total visit time\\
\begin{displaymath}
\MaxVisitWiFiCoI(t) = \left\{ \begin{array}{rl}
\CoIVisitCount_{\CoI_i}, & i = \arg \max_{\CoI \in \{ \AllCoIs^{\text{WiFi}} \cap \LocationContext_t \} }{\CoIVisitTime_\CoI} \wedge\\
& \LocationContext_t \cap \AllCoIs^{\text{WiFi}} \neq \emptyset \\
0, & \text{otherwise}\\
\end{array} \right .
\end{displaymath}
\noindent
\textbf{Device familiarity features.}\\
$\DevicesInContext$: Number of Bluetooth devices and familiar Bluetooth devices in device context $\DeviceContext_t$\\
\begin{displaymath}
\DevicesInContext(t) = \vert \DeviceContext_t \vert, \:\:
\FamDevicesInContext(t) = \vert \DeviceContext_t \cap \FamiliarDevices \vert\\
\end{displaymath}
$\AvgEncTimeFamDevices$: Average encounter time of familiar devices in $\DeviceContext_t$\\
\begin{displaymath}
\AvgEncTimeFamDevices(t) = \left\{ \begin{array}{rl}
\frac{\sum_{\Device \in \{ \DeviceContext_t \cap \FamiliarDevices \} }{\DeviceEncounterTime_\Device}}{\vert \DeviceContext_t \cap \FamiliarDevices \vert}, & \DeviceContext_t \cap \FamiliarDevices \neq \emptyset\\
0, & \text{otherwise}\\
\end{array} \right .
\end{displaymath}
$\AvgEncFamDevices$: Average number of encounters of familiar devices in $\DeviceContext_t$\\
\begin{displaymath}
\AvgEncFamDevices(t) = \left\{ \begin{array}{rl}
\frac{\sum_{\Device \in \{ \DeviceContext_t \cap \FamiliarDevices \} }{\DeviceEncounterCount_\Device}}{\vert \DeviceContext_t \cap \FamiliarDevices \vert}, & \DeviceContext_t \cap \FamiliarDevices \vert \neq \emptyset\\
0, & \text{otherwise}\\
\end{array} \right .
\end{displaymath}
The \ContextProfiler calculates context feature values based on a history of observation data and labels them based on user feedback. The feature values are used by the \ContextClassifier to train machine learning-based classifiers for classifying new observations. 
In the following sections, we show how we applied this context model on real-world context data to evaluate the effectiveness of the model and the \ContextClassifier.

%% file: implementation.tex
To evaluate the \ourname framework, 
we created a prototype implementation consisting of a \ContextDataCollector app, a \ContextProfiler and \ContextClassifier. The output of the \ContextClassifier was integrated with the \accesscontrollayer (cf. section \ref{accesscontrollayer}) to provide enforcement. 

\subsection{Data Collector}
\label{sec:ContextDataCollector}


For collecting context data, we implemented a \ContextDataCollector app for Android. It uses a background \Service to collect context data in intervals of 60 seconds. This is a required tradeoff between the battery lifetime and the quantity of collected data for reaching a battery lifetime of at least a working day (12h) on, e.g., the Samsung Galaxy Nexus and Nexus S devices. 
The collected data comprise location information, 
nearby Bluetooth devices and WiFi access points, acceleration sensor information as well as information about user presence and her interaction with apps \textsf{(Activities)}.%
\footnote{\ContextDataCollector is a generic solution collecting more data than required by the current \ContextProfiler.}

The \ContextDataCollector app also collects \emph{ground truth data}. The user regularly reports the perceived risk of device misuse in the current context by specifying the context to be ``safe'' (low risk of misuse) or ``unsafe'' (high risk of misuse). In addition, users are asked to classify the current context as ``home'' or ``work'', if the context has high privacy exposure due to context information being either \emph{private} or \emph{confidential}, respectively, or, ``public'' if the context has low privacy exposure. By using concise words that are easy to follow and relate to helped us in keeping the user interaction as simple as possible. 
To avoid misunderstandings, an introductory explanation was given to study participants beforehand.

\begin{figure}[htb]
  \centering
  \subfloat[Feedback using Context Feedback Buttons]{\includegraphics[scale=0.13]{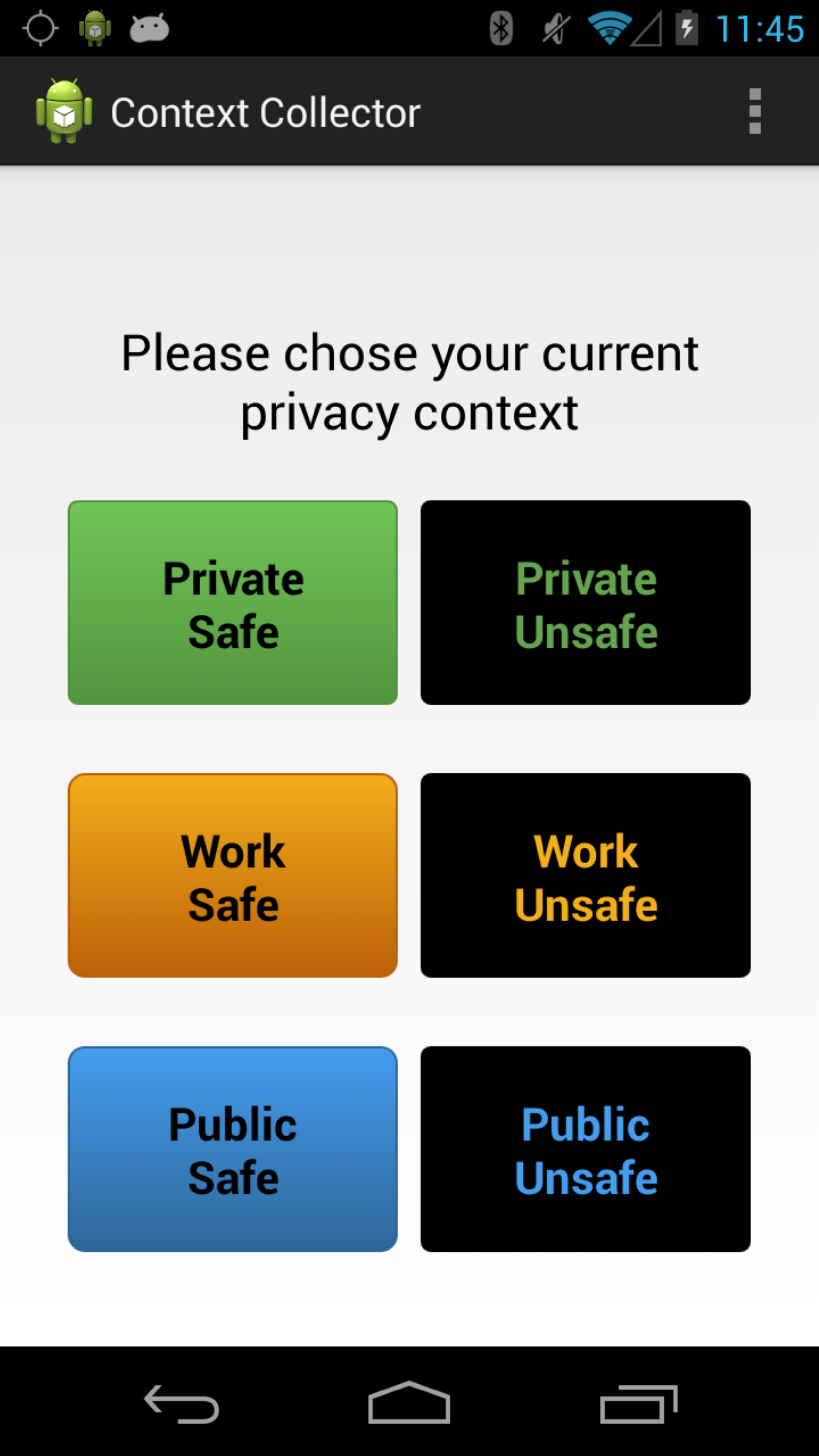}
  \label{fig:contextCollector1}
  }
  \quad
  \subfloat[Feedback using Context NFC Tags.]{\includegraphics[scale=0.13]{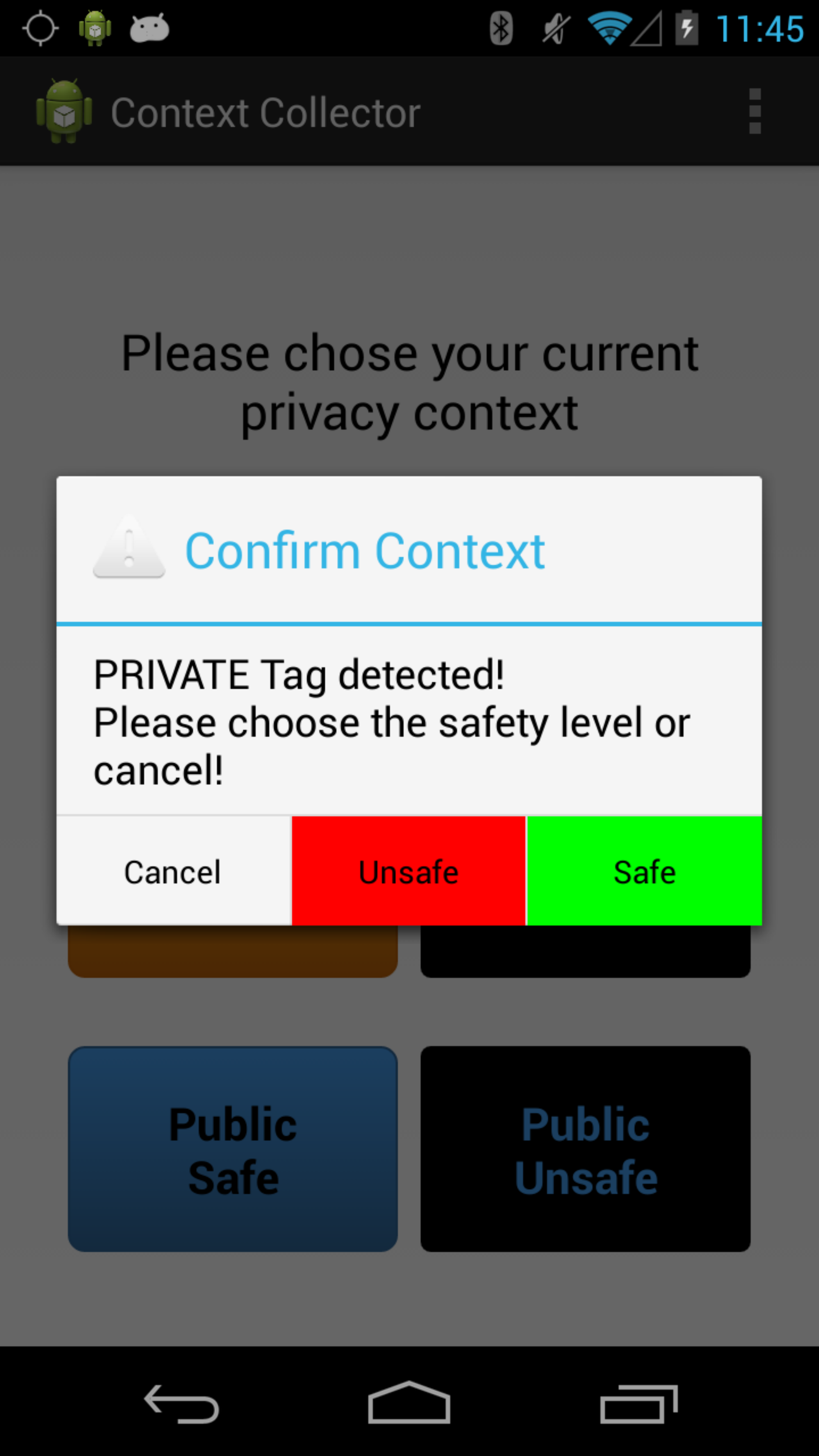}
  \label{fig:contextCollector2}
  }
  \caption{Android Data Collector App}
  \label{datacollectionapp}
\end{figure}

The users provided the above feedback for the current context either by using context feedback buttons on the device's UI or by using dedicated NFC tags provided to the participants for triggering context reporting (cf. Figure \ref{datacollectionapp}). The feedback UI was either spontaneously invoked by the user, or, if no ground truth had been provided during the last two hours, the app reminded the user to do so via sound, vibration and flashing LED notifications. Context and ground truth data were stored in a SQLite database and periodically uploaded to a server via HTTPS.

\subsection{Profiler and Classifier}

We implemented the functionality of the \ContextProfiler as off-line data processing scripts utilizing \texttt{bash} shell scripting, \texttt{awk} and \texttt{Python}. The scripts were used to identify individual GPS and WiFi CoIs for each user, and to calculate the familiarity of Bluetooth devices that the users had encountered during the data collection period. 
Scripts were also used for extracting the context feature vectors.

The functionality of the \ContextClassifier was realized and evaluated using the Weka data mining suite \cite{Hall2009} and its provided algorithm implementations for k-NN, Random Forest and Na\"ive Bayes classifiers.



%% file: evaluation.tex

To evaluate the context classification, the \ContextDataCollector app was installed on the Android smartphones of 15 test users having technical and non-technical backgrounds. A test user group of this size is large enough for verifying the validity of the concept and is in line with previous works evaluating context-aware access control by Riva et al.~\cite{riva2012progressive} ($n=9$) and Sadeh et al.~\cite{Sadeh2009} ($n=12$ and $n=19$). Users provided context and feedback data over a period of 68 days, 56 days per user on the average. The total dataset contained data from 844 distinct user days. On the average, users provided ground truth feedback on 46 days of the data collection period, resulting in a ground truth dataset containing $3757$ labeled data points.


From the collected data, the \ContextProfiler calculated personal context profiles and context features. The features were used by the \ContextClassifier to train classification models for predicting the privacy exposure and misuse risk levels of contexts. The context labels obtained through user feedback were used to attach class labels to the context feature vectors.

Each test user provided at least 50 or more feedback labels. This would roughly correspond to the user providing 2-3 feedbacks per day over a period of three weeks, which seems like a manageable burden on the user. After this initial training period of the \ContextClassifier, the need for explicit user feedback would significantly diminish. The user would need to provide only occasional corrective feedback in cases of incorrect predictions of the \ContextClassifier.

In constructing the \ContextClassifier, we experimented with three different machine learning algorithms:
1) A k-nearest neighbors (kNN) classifier, which
bases its prediction on comparing a testing datapoint to the $n$ closest observations to it in the training dataset. The prediction is the most frequent class label in this set of observations. 2) A Na\"ive Bayes (NB) classifier is a simple probabilistic classifier which has been successfully used, e.g., in spam e-mail detection~\cite{sahami1998bayesian}.
3) Random Forest (RF) is an ensemble method that is commonly used for classification tasks. It randomly picks subsets of input attributes and trains decision trees for them. It uses the most frequently predicted label provided by this set of tree classifiers as the final prediction. For each participant, we trained the \ContextClassifier using the labeled context feature vectors and evaluated the performance of the classifiers using 10-fold cross-validation. 

We assume that by default restrictive protection measures are in place (access to sensors disabled, device lock active). The \ContextClassifier's task is therefore to predict situations, in which the protections could be relaxed, i.e., if the context has low privacy exposure or a low risk of device misuse.

Even though most accurate results would be obtained by direct measurement of on-line enforcement on users' mobile devices, we had to rely on an offline evaluation of the {\ContextClassifier}s performance, since we wanted to be able to experiment with several different machine learning algorithms. Implementing or porting several different algorithms on the mobile device and conducting a separate user study for each of them was not feasible given the resource limitations. Therefore we intend to evaluate the performance of on-line enforcement in a subsequent user study involving devices with enhanced device locking functionality.

\begin{figure}[tbh]
\includegraphics[width=\columnwidth]{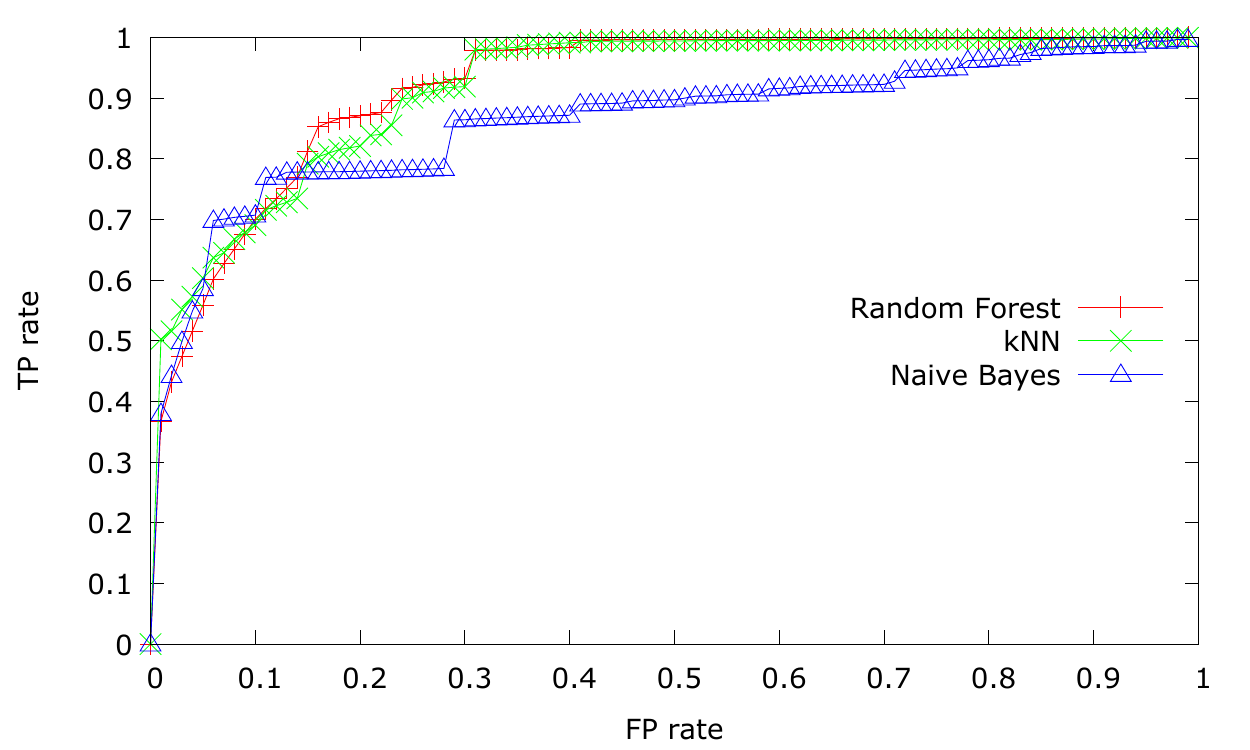}
\caption{Average receiver operating characteristic (ROC) curves for classifying contexts with low risk of device misuse.}
\label{fig:safe}
\end{figure}

%
%
%
\noindent
\textbf{Protecting against device misuse.}
Figure \ref{fig:safe} shows the average receiver operating characteristic (ROC) curves of the classifiers for users who provided at least five feedback datapoints for each context class.

All classifiers perform reasonably well on the testing data, providing usable results for practical use. The classifiers reach a true positive rate of approximately 70\% with a fairly moderate false positive rate of 10\%. This would mean that by applying a relaxed device locking scheme in low-risk contexts, we can potentially reduce the amount of unnecessary authentication prompts shown to the user by 70\%. Only one time in ten would a relaxed locking mechanism be enforced while the user is in a context with higher risk of misuse. This means that a thief or other unauthorized user would likely have a less than 10\% chance of finding the device in an unlocked state, when obtaining physical access to it. These results clearly outperform the progressive authentication scheme presented by Riva et al.~\cite{riva2012progressive}, who report a reduction of 42\% in unnecessary authentication prompts presented to the user.

\begin{figure}[tbh]
\includegraphics[width=\columnwidth]{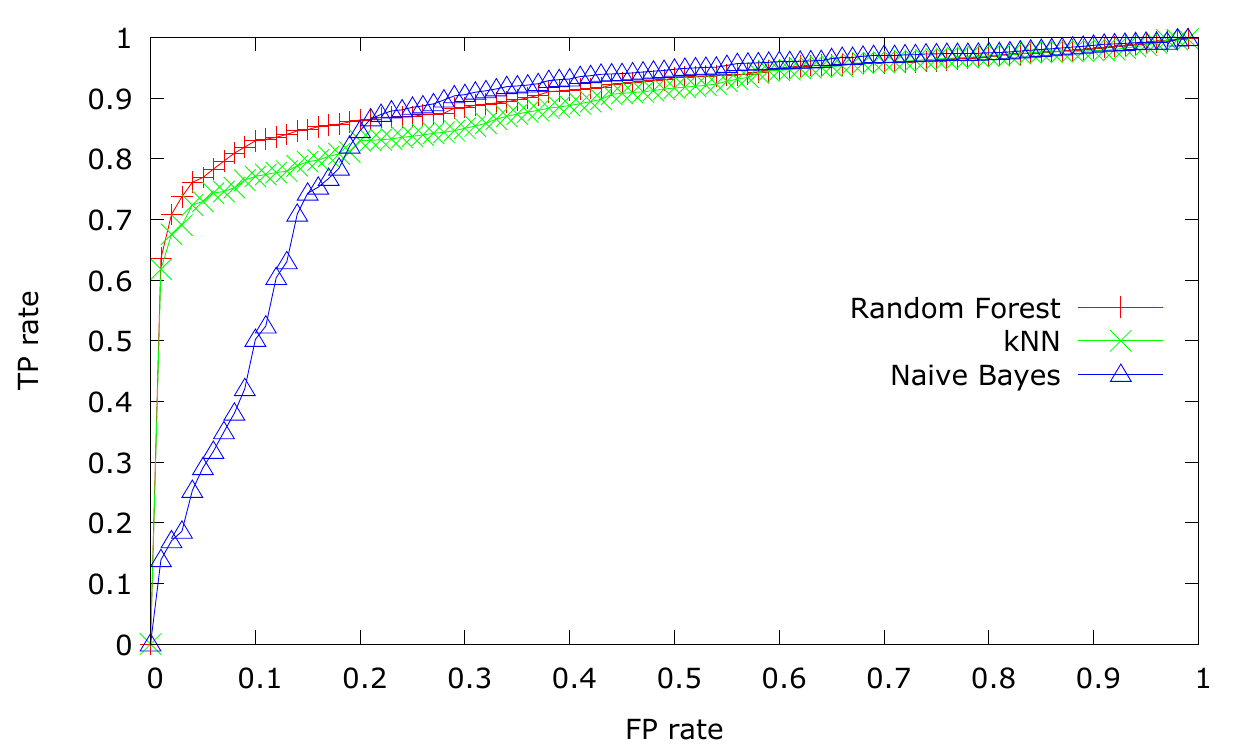}
\caption{Average ROC curves for classifying contexts with low privacy exposure.}
\label{fig:sensitive}
\end{figure}

\noindent
\textbf{Protecting against sensory malware.}
Figure \ref{fig:sensitive} shows the average performance of the classifiers in identifying contexts with low privacy exposure.


For this use case, the Random Forest and kNN classifiers provide best performance. They reach a true positive rate of 70\% at a very low false positive rate of 2-3.5\%. This would mean that if a sensory malware protection scheme with a 'default deny' policy is enforced, only in less than 3.5\% of the cases would access control be relaxed in contexts with high privacy exposure. In practice, this would severely limit a sensory malware application's ability to extract useful sensitive information about the user. 

Through the use of a default deny policy, our framework errs by default on the safe side, i.e., sensory malware is by default denied access to sensor information. The true positive rate of 70\% means that our scheme is able to relax the access restrictions to sensors in public or low-privacy exposure contexts in 70\% of the cases. The remaining 30\% can be handled through manual overriding of the default policy by the user. Fortunately, the use of context information by many apps is often user-driven, i.e., sensor data are utilized, when the user is actively using the app (e.g., using a navigation app to locate a nearby restaurant). Adding an override confirmation dialog to the user interaction in such situations should therefore be easy, since the device already is in the focus of the user's attention. This approach also has the benefit that the overriding action can be used as additional ground truth data for updating the classification model and thus improving subsequent classification accuracy.

%% file: accesscontrol.tex

\begin{figure}
\centering
\includegraphics[width=0.8\columnwidth]{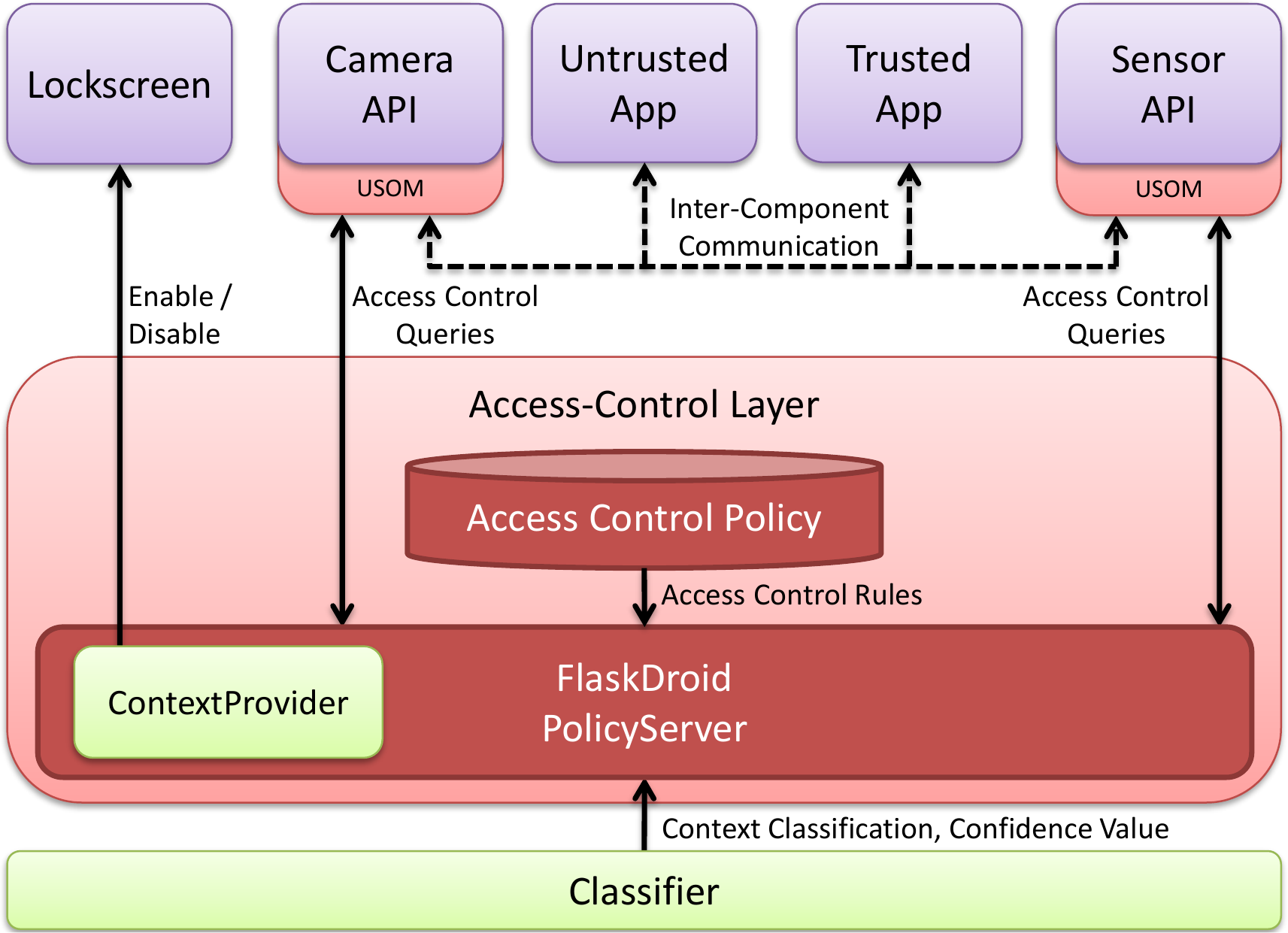}
\caption{Enforcement of Context-based policies}
\label{fig:AccessControlArchitectureDetails}
\end{figure}

To verify the applicability of our framework to practical access control enforcement, we integrated it with an \accesscontrollayer for which we adopted and adapted the \flaskdroid~\cite{flaskdroidUsenix} architecture, a fine-grained mandatory access control framework for Android 4.0.4 (cf. Figure~\ref{fig:AccessControlArchitectureDetails}). We now show how the combination of \ourname and \flaskdroid can address the previously defined use-cases, namely \emph{Resisting Sensory Malware} and \emph{Usable Device Lock} (cf. Section~\ref{usecases}). For our implementation we use a Samsung Galaxy Nexus smartphone.

\subsection{Implementation}
\flaskdroid extends \emph{Security Enhanced Android (SEAndroid)}~\cite{seandroid:ndss} with fine-grained type enforcement on Android's middleware layer. In \flaskdroid, Android components that provide access to sensitive resources, such as the \SensorService which provides access to sensor information, act as \userspaceobjectmanagers \textsf{(USOMs)} which control access to resources they manage. More specifically, \usoms control $operations$ from $subjects$ (i.e., apps) to $objects$ (e.g., data) using $types$ assigned to $subjects$ and $objects$. 



At boot time, \flaskdroids \PolicyServer (cf. Figure~\ref{fig:AccessControlArchitectureDetails}) parses an \AccessControlPolicy and proceeds to assign app types (e.g., \trusted or \untrusted) to all installed apps based on application metadata (e.g., package name or developer signature). Apps installed by the user are assigned types during their installation. Whenever apps request access to a \usom, for example the \SensorService to query the device's sensors or the \CameraService to take pictures, the \usom queries the \PolicyServer, which is part of Android's \SystemServer, for access control decisions. \flaskdroid supports conditional access control rules by means of \ContextProviders that evaluate the current context and enable or disable rules at runtime.



To meet our goals we extended \flaskdroid with additional \usoms and  
implemented a \ourname \ContextProvider. It uses the context classification information and confidence values provided by the \ContextClassifier to activate or deactivate conditional rules at runtime (cf. Figure~\ref{fig:AccessControlArchitectureDetails}) and to influence the \Lockscreen behaviour. The \ContextProvider can be tuned with individual user-, use-case and sensor-specific thresholds for the expected confidence values. These thresholds could be set, e.g., by specifying a desired maximal false positive rate and adjusting the confidence threshold accordingly based on the observed historical performance of the context classifier. Access to more sensitive context sensors like GPS could require a higher prediction confidence than less sensitive sensors like the magnetometer.

\noindent\textbf{Mitigation of Sensory Malware.} To mitigate, respectively reduce the effects of sensory malware (e.g., \emph{Place\-rai\-der}~\cite{Templeman2013} or SoundComber~\cite{Schlegel2011}), access control on the sensors of a device is required. For example, \emph{Placeraider} uses the device's camera and the acceleration sensor to covertly construct 3D images of the surroundings of the user. We transformed Android's \CameraService into an \usom which filters queries to the \textit{takePicture} and \textit{startPreviewMode} methods. Furthermore, we used \flaskdroid to filter acceleration sensor events delivered to \SensorEventListeners registered by apps. It should be noted that \flaskdroids original implementation of the \SensorManager \usom is insufficient to block sophisticated attacks, since the \SensorManager is loaded into the memory space of (potentially malicious) apps. Thus, we replaced \flaskdroids \SensorManager \usom with a corresponding \usom in Android's \SensorService, which is not under the control of apps.

Similarily, the combination of \ourname and \flaskdroid can address also other variants of sensory malware, such as \emph{Soundcomber}~\cite{Schlegel2011}, by identifying the relevant Android APIs, instrumenting them as \textsf{USOMs} and extending the \flaskdroid policy with corresponding conditional rules.

\noindent\textbf{Usable Device Lock.} 
To allow for changes in the Android \Lockscreen policy based on the current risk for device misuse, we use the \ourname \ContextProvider to configure Android's \Lockscreen dynamically at runtime. We modified Android's \Settings component to be notified by our \ContextProvider about changes in the current risk for device misuse by means of a \Broadcast \Intent. We further modified Android's \LockPatternKeyguardView which is used to display the Lockscreen to query the \Settings component for context information. While the device is used in a context with \textit{low} risk for device misuse, the \LockPatternKeyguardView class automatically dismisses the \Lockscreen. Whenever the device is rebooted or the risk for device misuse changes to \textit{high}, a low-watermark mechanism ensures that the \Lockscreen is always displayed regardless of the current risk for device misuse. This mechanism is required to prevent an attacker from bypassing the \Lockscreen by changing the context, emulating a context the user considers to have low risk for device misuse or rebooting the device. In addition, to mitigate the effect of sensory malware which uses the acceleration sensor as a side channel to derive user credentials (e.g., Lockscreen PIN or password)~\cite{Xu:2012:TIU:2185448.2185465,Owusu:2012:API:2162081.2162095,Cai:2011:TIK:2028040.2028049}, we use the \SensorService \usom, our \ContextProvider and corresponding conditional access control rules to block access to the acceleration sensor by \thirdparty apps while the \Lockscreen is displayed.


\subsection{Evaluation}

\noindent\textbf{Mitigation of Sensory Malware.} 
To mitigate the effects of the \emph{PlaceRaider}~\cite{Templeman2013} sensory malware we designed a \flaskdroid policy to assign the type \emph{trusted} to all pre-installed system apps (e.g., the camera app), and the type \emph{untrusted} to all \thirdparty apps. In a real-world scenario this trust level could be derived from the app's reputation in an app market. We use conditional access control rules for the \CameraService and \SensorService \usoms to prevent all \emph{untrusted} apps from accessing the acceleration sensor and the camera when the risk for privacy exposure is \textit{high}.


We tested our implementation using a slightly modified version of the \emph{PlaceRaider} malware generously provided to us by its authors\footnote{The sample we received is incompatible with Android 4.0.4.}. By installing the malware on our device and logging the context information and access control decisions we verified that \flaskdroid successfully filtered all data delivered from Android's \SensorService and \CameraService components to the \untrusted \emph{PlaceRaider} app when the risk for privacy exposure was \textit{high}, thus rendering the attack futile. We further verified that \trusted apps could still use the sensors and the camera. No false positives or false negatives emerged during the evaluation of the \accesscontrollayer, which is not surprising since it merely enforces context-dependent access control rules.

To evaluate the performance impact of the \accesscontrollayer we implemented an app which automatically triggers $10,000$ access control queries by reading sensor data and taking pictures. On average, the \accesscontrollayer caused an overhead $\mu$ of 
4.9 ms (standard deviation $\sigma$ 17.6 ms) for the \SensorService and \CameraService \usoms on a Samsung Galaxy Nexus smartphone.
The high standard deviation $\sigma$ 
is caused by the garbage collector used in Android's Dalvik Virtual Machine: While studying Android's system logs we noticed that during the irregularly slow access control queries, which are responsible for the high standard deviation, the garbage collector started and caused a stall. Overall, 95\% of all access control decisions are handled in less than 4.2 ms, which we consider reasonable.


\noindent\textbf{Usable Device Lock.}
To test our implementation of the context-aware device lockscreen we modified the Android operating system to periodically wake the device from sleep and switch on the screen. We furthermore installed a synthetic malware, which registers \SensorEventListeners in Android's \SensorService to be notified of acceleration sensor readings. By logging and analyzing the \Lockscreen behavior, context information and sensor readings we verified that the \Lockscreen was only automatically dismissed in valid situations and that our synthetic malware did not receive any sensor readings while the \Lockscreen was active.

%% file: relatedwork.tex

In the digital society, context data have been ex\-ten\-sive\-ly used to analyze numerous aspects of human everyday life. Examples range from the prediction of health status by interpreting context data~\cite{Madan2010} to analyzing ethnographics~\cite{Hoflich2006} or person matching based on similar interests~\cite{Eagle2005}. Our framework brings this idea of contextual analysis to the area of security and privacy protection for the most important tool of modern life - the smartphone.

A number of works have approached the problem of con\-text-aware access control. Contrary to our work, all of them rely on user-defined or pre-defined policies in the form of role definitions, conditions on context parameters, or context-dependent rules. For example, 
Covington et al.~\cite{Covington2002} use a \emph{Generalised Role Based Access Control} (GRBAC) model
utilizing Environment Roles that are activated and deactivated based on context observations, and
 Damiani et al.~\cite{Damiani2007} utilize roles in their spatially-aware RBAC model
using location as a component for access control decisions.

Others have used user- or pre-defined policies conditioned on context parameters. Examples include
Sadeh et al.~\cite{Sadeh2009} who investigate a policy definition and management system for the \emph{PeopleFinder} application and 
Kelley et al.~\cite{Kelley2008}, who introduced a user-controllable policy learning system that builds on incremental policy improvements proposed to the users based on recorded history events.
For mobile devices, Bai et al. propose a solution for fine-grained usage control on Android~\cite{CONUCON}. Their work extends the UCON access control model~\cite{UCON} by using context information (e.g., location and time) as an additional input for policy decisions.

Hull et al.~\cite{Hull2004} present the \emph{Houdini} framework 
for mitigating the complexity that value-based customization of policies implies by using user-provided higher-level preferences to generate rules for privacy enforcement. They mention the possibility for automatically-learned preferences, 
but do not provide support for such automation at the time of writing.

Many recent papers have addressed context-aware access control enforcement on mobile devices. For example, 
Conti et al.~\cite{Conti2012} describe the \emph{CRePe} framework for Android for enforcement of context-dependent access control policies 
allowing or denying access to specific resources depending on the currently detected active context.
In the \emph{MOSES} framework~\cite{Russello2012} Rusello et al. propose a combination of dynamic taint tracking using the \emph{TaintDroid} architecture~\cite{Taintdroid} and policy enforcement on Android's middleware layer to enable context based access control on resources and apps with the goal of providing isolated environments called security profiles. Similarly, the \emph{TrustDroid}~\cite{TrustdroidSPSM} architecture provides lightweight security domain isolation on Android with basic support for context-based network access control policies. Saint~\cite{Ongtang09semanticallyrich} features a context-aware fine-grained access control framework for Android, which focuses on enabling app developers to define context-dependent runtime constraints on inter-app communication. Nauman et al. present \emph{Apex}~\cite{Nauman:apex}, which extends the Android operating system with conditional permissions. It provides to some extent support for context-based access control by allowing the user to define context-dependent resource restrictions (e.g., based on the time of day). All of these works heavily rely on user- or pre-defined rules, whereas our work relies on dynamic context classification utilizing machine learning as a source for access control enforcement.
Also, in contrast to \emph{MOSES}, \emph{TrustDroid} and \emph{Apex}, our access control architecture is based on the more generic and flexible \emph{FlaskDroid} platform~\cite{flaskdroidUsenix}, which is also able to cover (most of) the use cases described in \emph{Saint}.

A recent patent application by Bell et al.~\cite{Bell2012} discloses a system using context-triggered policies controlling the access of applications to sensors and other resources on a smartphone. Also their approach relies on either pre-defined policies or policies uploaded to the devices by external entities.

Addressing the problem of more usable user authentication on mobile devices,
Riva et al.~\cite{riva2012progressive} use various contextual clues to (partially) authenticate the user by
estimating the likelihood that the user is in
proximity and use this information to configure the device lock. 
Similarly, Hayashi et al.~\cite{Hayashi:CASA} introduce \emph{Context-Aware Scalable Authentication}, an approach which uses the location of the device
in a probabilistic framework to determine the active authentication factors to be used for user authentication (e.g., PIN or password) on smartphones. Although we cover a similar use case as these papers, our approach is very different. We do not authenticate the user, but rather adjust device locking behavior based on automatic classification of the context according to its perceived risk level.



Kang et al.~\cite{Kang2005} introduced the idea of time-based
clustering of position observations, which Zheng
et al.~\cite{Zheng2010} used to introduce the concepts of stay points and stay regions, further developed by Montoliu
et al.~\cite{Montoliu2013}. We adopt a slightly modified form of the notion of stay regions to define our GPS-based CoIs. In addition, we also extend the notion of a stay points to non-locational data in the form of WiFi stay points.
Dousse et al.~\cite{Dousse2012} have successfully demonstrated the use of WiFi fingerprints for identifying and detecting places based on WiFi. We adopt a simplified version of their place identification scheme considering only intersections of WiFi snapshots for our WiFi-based CoI detection.

Gupta et al.~\cite{DBLP:conf/socialcom/0002MAN12} were the first to use
context profiling and the notion of CoI and device familiarity for estimating the 'safety' level of a context. Their system relied on a simple heuristic model based on time-discounted familiarity measures and suffered from having to specify a fixed threshold for distinguishing between context classes, which fails to take into account context- and user specific differences. Since we apply a sophisticated context model and more powerful machine learning models for context classification, our approach is capable to take better into account also context- and user-specific differences in perceptions of risk level and privacy exposure.


%% file: conclusions.tex
In this paper, we described \ourname, a context-aware access control framework for mobile devices utilizing automated  classification of contexts based on sensed context data. We
applied it to two concrete smart\-phone-related use cases:
defending against sensory malware and device misuse.
We showed that context classification can be used for context-aware access control enforcement, effectively addressing true security concerns that smartphone users have.
In this, however, we do not see the task merely as a prediction problem, but rather we consider true contextuality as a continuous process of learning from and adapting to the individual needs and preferences of smartphone users.

Having validated the effectiveness of \ourname, the next step is to evaluate its usability.  We plan to
  implement on-device versions of the \ContextProfiler and \ContextClassifier and create a mobile app for user studies focusing on the usability aspects related to our framework. We intend also to develop further richer context models incorporating more context sensors, and addressing other context-aware access control use cases.
